\documentclass[%
 reprint,
 amsmath,amssymb,
 aps,superscriptaddress,
]{revtex4-1}

\usepackage{color} 
\usepackage{amsmath}
\usepackage{natbib}
\usepackage[export]{adjustbox}
\usepackage{graphicx}
\usepackage{dcolumn}
\usepackage{bm}
\usepackage{subfigure}
\usepackage{tikz}
\usetikzlibrary{arrows}

\tikzset{font={\fontsize{5pt}{5}\selectfont}}

\begin{document}


\title{Angle-resolved non-resonant two-photon single ionization of argon using 9.3 eV photons produced via high harmonic generation}

\author{Kirk A. Larsen}
\email{klarsen@lbl.gov}
\affiliation{%
 Graduate Group in Applied Science and Technology, University of California, Berkeley, CA 94720, USA}
\affiliation{%
 Chemical Sciences Division, Lawrence Berkeley National Laboratory, Berkeley, CA 94720, USA}%
 
\author{Daniel S. Slaughter}
\affiliation{%
 Chemical Sciences Division, Lawrence Berkeley National Laboratory, Berkeley, CA 94720, USA}%
 
\author{Thorsten Weber}
\affiliation{%
 Chemical Sciences Division, Lawrence Berkeley National Laboratory, Berkeley, CA 94720, USA}%

\date{\today}

\begin{abstract}
We present an experimental study on the photoionization dynamics of non-resonant one-color two-photon single valence ionization of neutral argon atoms. Using 9.3 eV photons produced via high harmonic generation and a 3-D momentum imaging spectrometer, we detect the photoelectrons and ions produced from non-resonant two-photon ionization in coincidence. Photoionization from the $3p$ orbital produces a photoelectron scattering wave function with $p$ and $f$ partial wave components, which interfere and result in a photoelectron angular distribution with peak amplitude perpendicular to the VUV polarization. The comparison between the present results and two previous sets of theoretical calculations [Pan, C. \& Starace, A. F. (1991). \textit{Physical Review A}, 44(1), 324., and Moccia, R., Rahman, N. K., \& Rizzo, A. (1983). \textit{Journal of Physics B: Atomic and Molecular Physics}, 16(15), 2737.] indicates that electron-electron correlation contributes appreciably to the two-photon ionization dynamics.
\end{abstract}

\pacs{Valid PACS appear here}
\maketitle



The photoionization dynamics of multi-electron atomic and molecular systems are influenced by electron-electron correlation. Non-resonant two-photon ionization can probe such correlation effects in both the initial and final states of the target. A particularly sensitive observable is the photoelectron angular distribution (PAD), which can provide a detailed view of the underlying mechanisms involved in the photoionization process and its correlated nature, e.g. information on the role of continuum states and interchannel coupling~\cite{Boll,Yip,Lucchese0,Lucchese1,Lucchese2,Lucchese3,Rescigno,Codling,Altun,Dehmer,Lin,Manson}. The PAD emerges from a coherent summation over a set of final continuum states. The sensitivity of the PAD to electron-electron correlation arises from its dependence on the amplitudes and phases of the different partial wave components of the coherent sum. These distinct angular momentum components can interfere to create nodes and antinodes in the PAD. 

PADs are uniquely characterized by their energy-dependent anisotropy parameters, or $\beta$ parameters. The number of $\beta$ terms used to describe the PAD increases with the photon order. As such, a two-photon PAD can exhibit more anisotropy and structure than the corresponding one-photon PAD. Previous two-photon investigations of the anisotropy parameters in neon and argon have been realized using two-color two-photon above-threshold ionization schemes \cite{Leone,Mondal,Dusterer}, where ionization was performed with a VUV field in the presence of a strong NIR dressing field that generated photoelectron sidebands. This can make comparison with theory very challenging. Measurements that lie within the perturbative limit and target non-resonant bound-continuum transitions, driven by the second photon (rather than continuum-continuum transitions), are highly sensitive to electron-electron correlation and can be achieved in a one-color two-photon ionization scheme by exclusively using a VUV field with a photon energy in a non-resonant region below the ionization threshold. However, measuring a PAD from \textbf{n}on-resonant \textbf{o}ne-color \textbf{t}wo-\textbf{p}hoton \textbf{s}ingle \textbf{i}onization (NOTPSI) in an atomic gas requires sufficiently high VUV intensities to enable nonlinear processes. Since high intensity ultrashort VUV light sources are limited to a small number of free electron lasers (FELs) and tabletop high-order harmonic generation (HHG) systems, angle-resolved measurements on NOTPSI in rare gases are scarce.

Over the years, several studies have investigated one-color two-photon ionization in rare gases, first using HHG based light sources \cite{Miyamoto,Sekikawa3}, and later using VUV FELs \cite{Sato,Meyer,Ma}. PADs were measured in helium at several photon energies, across both the resonant and non-resonant regions, in Ref. \cite{Ma}. Here, anisotropy parameters as well as amplitude ratios and phase differences of the partial wave components of the scattering wave function could be extracted, due to the simple nature of the target. By moving to more complex many-electron systems, more terms and higher angular momentum components contribute to the photoelectron scattering wave function, and many-electron effects become more significant. This increase in complexity represents a great challenge for experiment and theory alike.

Previous theoretical studies on angle-resolved two-photon ionization in helium have indicated that a single-active-electron picture appears to be a valid approach in describing the photoionization dynamics for photon energies below the ionization threshold (and even in the above threshold region) \cite{Boll}. It is unlikely that this is true for more complex core targets. This compels angle-resolved measurements in more complicated systems, where many active and correlated electrons are required to describe the photoionization dynamics. To our knowledge, no angle-resolved measurements exist for complex multi-electron systems such as argon, were electron-electron correlation is expected to play a more significant role in the photoionization dynamics than in simple systems like helium. The aim of this experimental investigation is to reveal, for the first time, clear contributions from electron-electron correlation in the PADs emerging from NOTPSI of argon.

Despite the paucity of experimental data, the problem has not escaped theoretical treatment. Over a quarter of a century ago, the $\beta$ parameters for one-color two-photon single ionization (including NOTPSI) were calculated for argon using a Hartree-Fock approach \cite{Starace3} providing uncorrelated and Coulomb correlated results, and a random phase approximation calculation \cite{Moccia}, which neglected electron-electron correlation. The uncorrelated results of Ref.~\cite{Starace3,Moccia} are somewhat ambiguous due to discrepancies between the calculations performed in the length and velocity gauge at various photon energies. The correlated results of Ref.~\cite{Starace3} show better gauge invariance in both the resonant and non-resonant two-photon ionization regions and the computed $\beta$ parameters suggest maximum photoelectron emission perpendicular to the ionizing field at 9.3~eV. However, to our knowledge, these calculations have for decades remained unverified by any experimental measurement.

In this work, we present results on angle-resolved NOTPSI of argon from the $3p$ orbital using 3-D momentum imaging, where the photoelectron and ion are measured in coincidence. Using a 400 nm driving field, we produce and select VUV photons with an energy of 9.3 eV via HHG, which are then used to perform NOTPSI. Interference between different angular momentum components of the photoelectron wave function results in a PAD exhibiting maximum intensity perpendicular to the ionizing VUV field. These experimental results are compared against previous calculations, which suggest that electron-electron correlation considerably influences the photoionization dynamics.


The valence photoionization dynamics in neutral argon were investigated using the cold target recoil ion momentum imaging~\cite{dornerColdTargetRecoil2000,Ullrich,Jahnke,Sturm} (COLTRIMS) technique. Here, the photoelectron-ion pair produced by NOTPSI are collected with full 4$\pi$ solid angle, and their 3-D momenta are measured in coincidence, on an event-by-event basis. The charged particles are guided by parallel DC electric and magnetic fields (15.55 V/cm, 3.72 G) towards position- and time-sensitive detectors at opposite ends of the spectrometer. The detectors consist of a multi-channel plate (MCP) chevron stack with a delay-line anode readout~\cite{Roentdek,Jagutzki}. The electron and ion detectors are a three layer hex-anode with a 80 mm MCP, and a two layer quad-anode with a 120 mm MCP, respectively. The 3-D momentum of each charge carrier is encoded into its hit position on the detector and its time-of-flight relative to the laser trigger.

The laser system has been described previously \cite{Sturm}, but we briefly highlight a few modifications made to the system below. A Ti:sapphire near-infrared (NIR) laser system produces 12 mJ, 45 fs pulses at 50 Hz, which are frequency doubled using a 0.25 mm thick beta-barium borate (BBO) crystal, where the copropagating 800~nm NIR and 400~nm blue fields are then separated using two dichroic mirrors. The reflected blue photons ($\sim$3.6 mJ, $\sim$50 fs) are focused (f = 6~m) into a 10 cm long gas cell containing 3 Torr of krypton to generate VUV odd harmonics via HHG. The resulting VUV frequency comb is then separated from the 400 nm fundamental by reflection from three silicon mirrors near Brewster's angle for the 400~nm field, resulting in a suppression of the fundamental by a factor of $<10^{-6}$. The 3\textsuperscript{rd} harmonic (133 nm, 9.3 eV) is isolated by transmission through a 0.25 mm thick MgF$_{2}$ window, which totally suppresses the 5$^{th}$ harmonic and above. The femtosecond pulse duration of the 3\textsuperscript{rd} harmonic is also maintained, while the residual 400 nm pulse is temporally separated from the 3\textsuperscript{rd} harmonic pulse by $\sim$700 fs, due to the difference in the group velocity dispersion (GVD) of the window at $\omega_0$ and $3\omega_0$ \cite{Allison,Li}. After transmission through the window, we estimate the pulse duration of the 3\textsuperscript{rd} harmonic to be $\sim$30 fs, based on its spectral bandwidth, its estimated attochirp, and the thickness and GVD of the MgF$_{2}$ window \cite{Sekikawa1,Sekikawa2}. The femtosecond 9.3 eV pulses are then back-focused (f = 15 cm) into the 3-D momentum imaging spectrometer using a protected aluminium mirror, the reflectance of which has been measured to be 43\% at 9.3 eV \cite{Larsen}. The pulse energy of the 3\textsuperscript{rd} harmonic on target is approximately 10 nJ, which was measured using a pair of broadband VUV filters (Acton Optics FB130-B-1D.3) and a calibrated photodiode. 

A beam of argon atoms is prepared from an adiabatic expansion through a 0.03 mm nozzle, which is then collimated by a pair of skimmers. This atomic jet propagates perpendicular to the focusing VUV beam, where the two intersect in the interaction region  (approximately 0.01 $\times$ 0.01 $\times$ 0.20 mm) of the spectrometer, resulting in a NOTPSI rate of $\sim$0.3 events per VUV pulse. 


The ground state electronic configuration of argon is $1s^2 2s^2 2p^6 3s^2 3p^6$ ($^1S$). Ionization from the $3p$ orbital results in the ground electronic state of the cation Ar$^+$, a $^2P$ state. From two-photon selection rules, the final states must have either $^1S$ or $^1D$ total symmetry, while the photoelectron wave function must be either a $p$- or an $f$-wave. It follows that we can express the allowed final states in the three forms listed below:

\begin{equation}
    ^1S: ~3p^5~^2P + \epsilon p
\label{1S}
\end{equation}
\begin{equation}
    ^1D: ~3p^5~^2P + \epsilon p
\label{1Dp}
\end{equation}
\begin{equation}
    ^1D: ~3p^5~^2P + \epsilon f
\label{1Df}
\end{equation}

In (\ref{1Dp}) and (\ref{1Df}) above, we see that the $^1D$ final state contains contributions from two different photoelectron angular momentum components, $p$- and $f$-waves. The coherent sum of these two partial waves can create an interference pattern in the PAD. Since the initial state has total magnetic quantum number $M=0$, so too must the final states. Hence the $m$ value of the photoelectron and ion wave functions must sum to 0. From this restriction, we see that only $m=0,\pm1$ values of the $f$-wave component can contribute, while all $m$ values of the $p$-waves may contribute. These photoelectron states are paired to states of the core with the appropriate $m$ value.

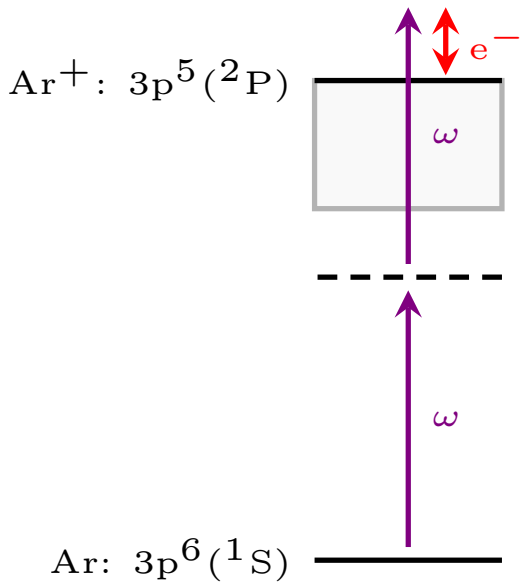
\begin{figure}[h!]
\centerline{
  \resizebox{8cm}{!}{
    \begin{tikzpicture}[
      scale=0.5,
      level/.style={thick},
      virtual/.style={thick,densely dashed},
      trans/.style={thick,->,shorten >=2pt,shorten <=2pt,>=stealth,color=violet},
      electron/.style={thick,<->,shorten >=2pt,shorten <=2pt,>=stealth,color=red}
    ]
    \filldraw[color=gray!60, fill=gray!5, thick](2cm,11.55em) rectangle (0cm,15.76em);
    \draw[level] (2cm,0em) -- (0cm,0em) node[left] {Ar: 3p$^6$($^1$S)};
    \draw[level] (2cm,15.76em) -- (0cm,15.76em) node[left] {Ar$^+$: 3p$^5$($^2$P)};
    \draw[virtual] (2cm,9.3em) -- (0cm,9.3em) node[midway,above]{};
    \draw[trans] (1cm,0em) -- (1cm,9.3em) node[midway,right] {$\omega$};
    \draw[trans] (1cm,9.3em) -- (1cm,18.6em) node[midway,right] {$\omega$};
    \draw[electron] (1.4cm,15.5em) -- (1.4cm,18.6em) node[midway,right] {e$^-$};
    \end{tikzpicture}
  }
}
\caption{An energy level diagram depicting the NOTPSI pathway from the $3p$ orbital of Ar at 9.3 eV. The grey box indicates the region containing bound excited states, the first appearing at 11.55 eV. The ionization potential of Ar is 15.76 eV, hence the red double-arrow corresponds with a photoelectron kinetic energy of 2.84 eV.}
\label{fig:LevelDia}
\end{figure}

A diagram depicting the NOTPSI pathway in the present experiment is shown in Fig~\ref{fig:LevelDia}. The grey box indicates the region containing the bound excited states of argon, beginning at 11.55~eV. The ionization potential of argon is 15.76~eV, while the two-photon energy is $\sim$18.6~eV, which results in an expected photoelectron kinetic energy of roughly 2.8~eV. Ionization of the ground state atoms via non-resonant two-photon absorption populates an ionic state and releases an electron into the continuum, with allowed final states listed in (\ref{1S})-(\ref{1Df}) above.

\begin{figure}[h!]
    {
    \subfigure[]{%
        \includegraphics[width=8.0cm]{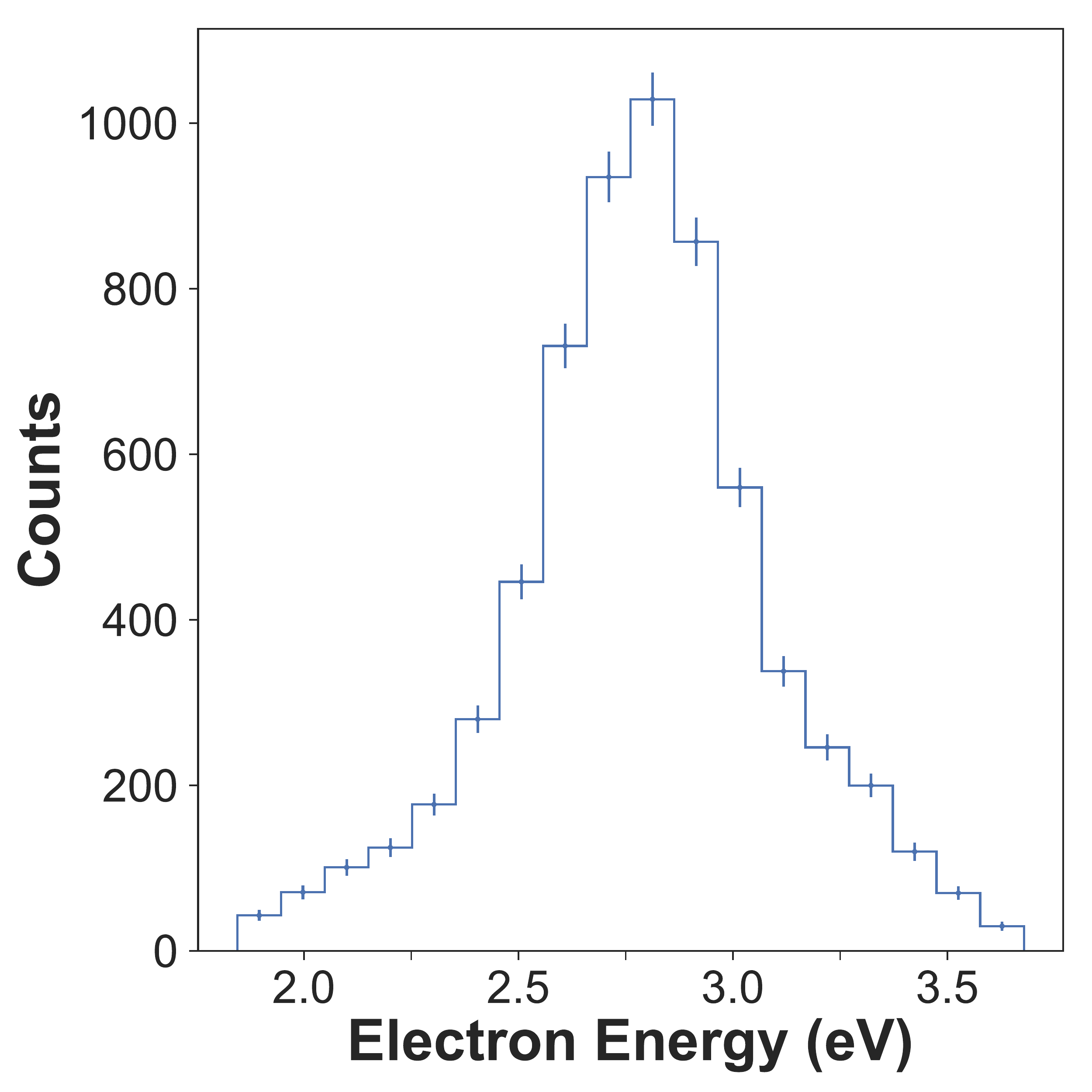}}}
    {
    \subfigure[]{%
        \includegraphics[width=8.0cm]{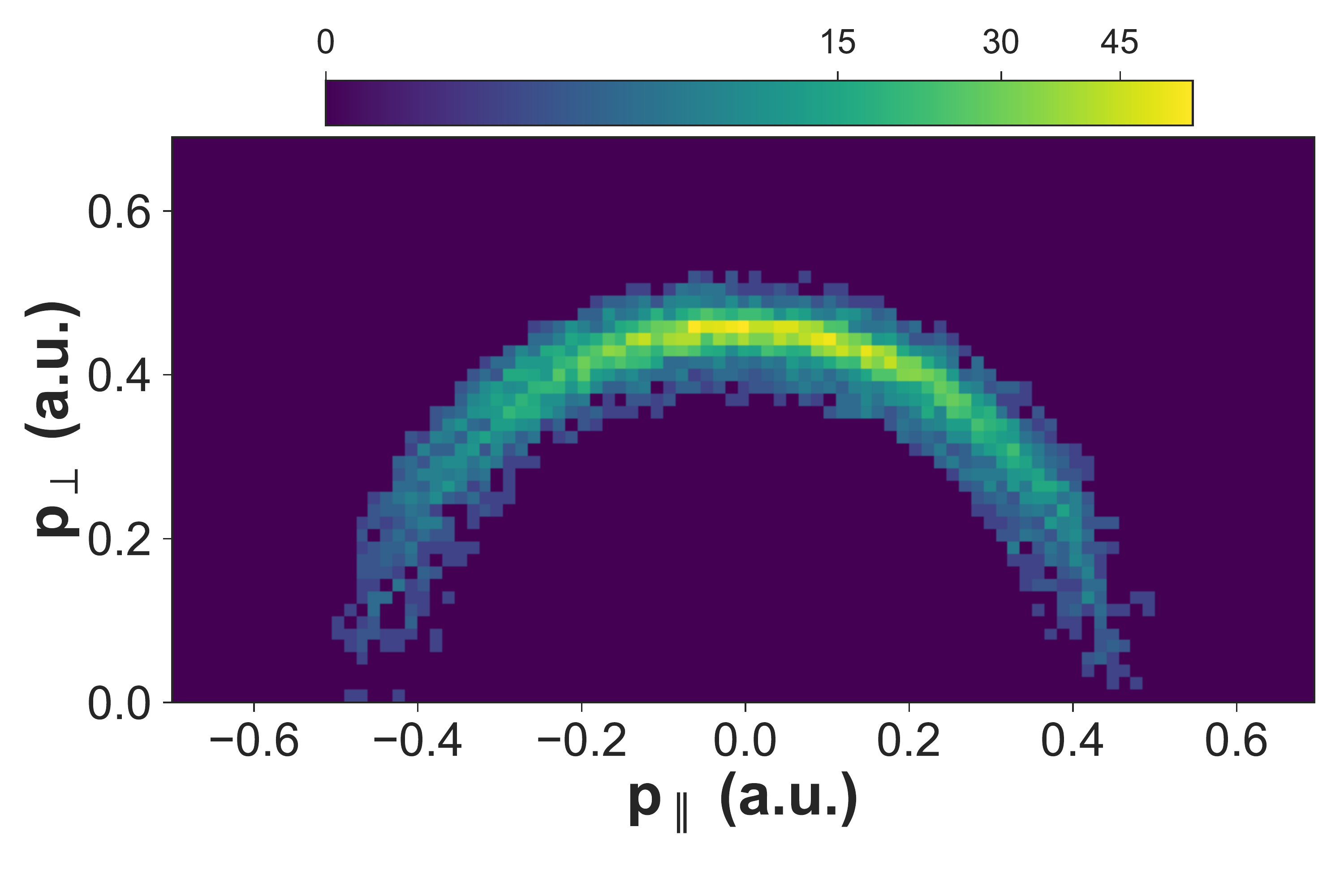}}}
\caption{(a) The photoelectron energy spectrum and (b) momentum distribution parallel versus perpendicular to the VUV polarization.}
\label{fig:Ee_p}
\end{figure}

The measured photoelectron kinetic energy spectrum is presented in Fig~\ref{fig:Ee_p} (a). Here, we observe a single peak centered at 2.8 eV, with a full width at half maximum (FWHM) of $\sim$400 meV, indicative of the two-photon spectral bandwidth of the 3\textsuperscript{rd} harmonic (convolved with the electron energy resolution of the spectrometer). The photoelectron momentum distribution transverse versus parallel to the VUV polarization vector is shown in Fig~\ref{fig:Ee_p} (b), where we observe electron emission peaking towards high transverse momentum and low longitudinal momentum. To gain more insight into the photoelectron emission pattern, we turn to the angle-differential photoionization cross section.

For two-photon ionization of a target atom by linearly polarized light, the angle-differential photoionization cross section is given by

\begin{equation}
    \frac{d\sigma}{d\Omega} = \frac{\sigma_0}{4\pi} [1 + \beta_{2}P_{2}(\cos\theta) + \beta_{4}P_{4}(\cos\theta)]
\label{PIC}
\end{equation}

where $\sigma_0$ is the total photoionization cross section, $\theta$ is the angle between the photoelectron momentum vector and the polarization vector of the light, $\beta_{2}$ and $\beta_{4}$ are the second and fourth order anisotropy parameters, and $P_{2}$ and $P_{4}$ are the second and fourth order Legendre polynomials in variable $\cos\theta$ \cite{Reid}. The measured angle-differential photoionization amplitude is presented in Fig 3. Equation~\ref{PIC} has been applied to fit the data (solid red line) using the projection method discussed in \cite{Liu}, where the error on the $\beta$ parameters is determined via statistical bootstrapping \cite{Efron}. The $\beta$ parameters retrieved from the fit are $\beta_{2}$ = -0.93$\pm$0.02, $\beta_{4}$ = 0.25$\pm$0.03. 

\begin{figure}[h!]
\includegraphics[width=8.0cm]{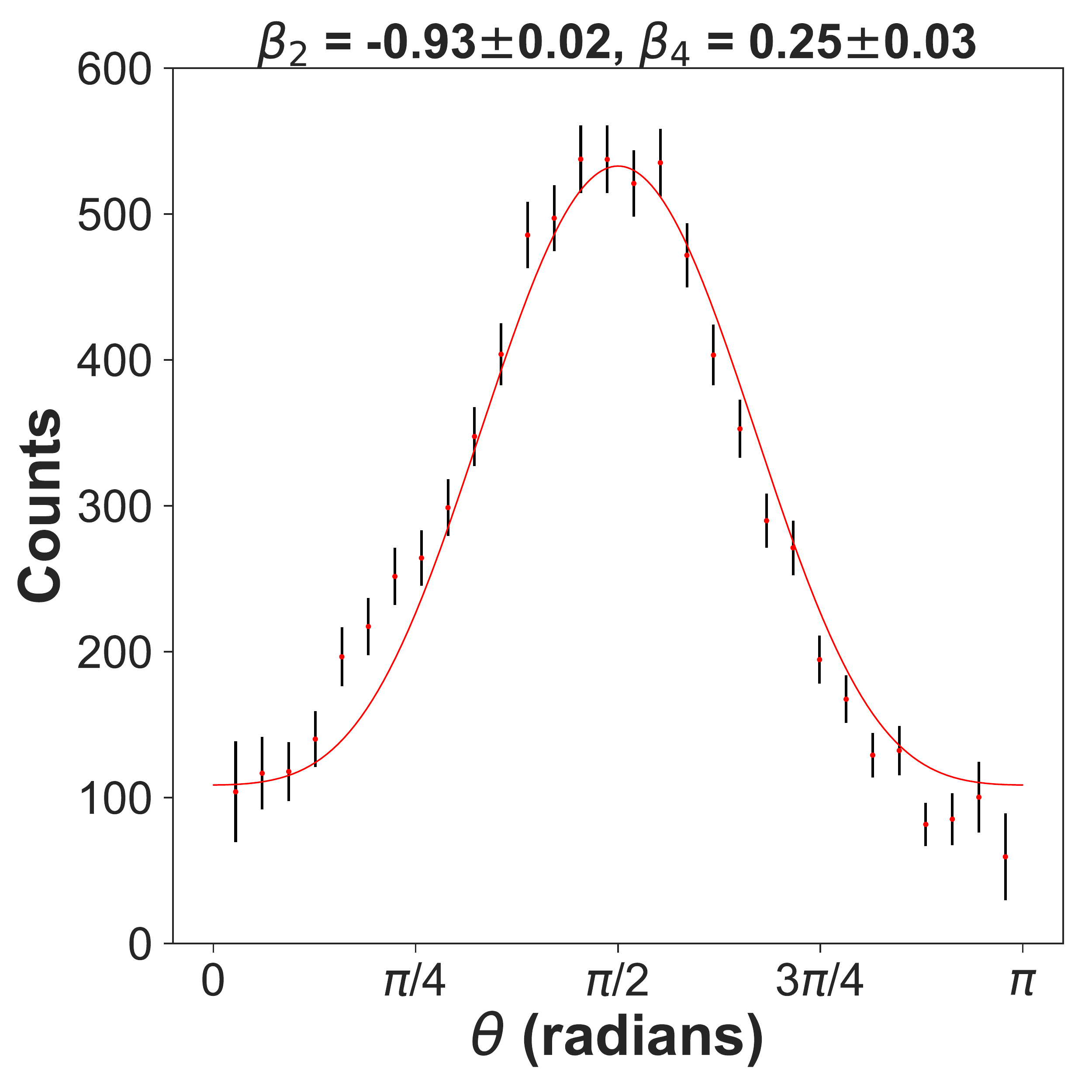}
\caption{The angle-differential photoionization cross section for NOTPSI of Ar at 9.3 eV. The experimental data is fit using Equation~\ref{PIC}, where the retrieved $\beta$ parameters are displayed above the plot.}
\end{figure}

The PAD exhibits peak intensity at angles near $\pi/2$, corresponding with maximum photoelectron emission perpendicular to the VUV polarization. Intensity minima occur along the VUV polarization direction, near 0 and $\pi$. We attribute these features to the interference between the different $p$- and $f$-wave components of the photoelectron scattering wave function. These two angular momentum components destructively interfere along the polarization direction, yielding an angle-differential amplitude that peaks perpendicular to the field. This interference is analogous to the interference between the photoelectron $s$ and $d$ partial waves in photodetachment of I$^-$ and O$^-$ ~\cite{Mabbs,Hall,Zare}. 

\begin{figure}[h!]
    {
    \subfigure[]{%
        \includegraphics[width=7.0cm, trim= 0.5cm 3.9cm 0.5cm 4cm, clip]{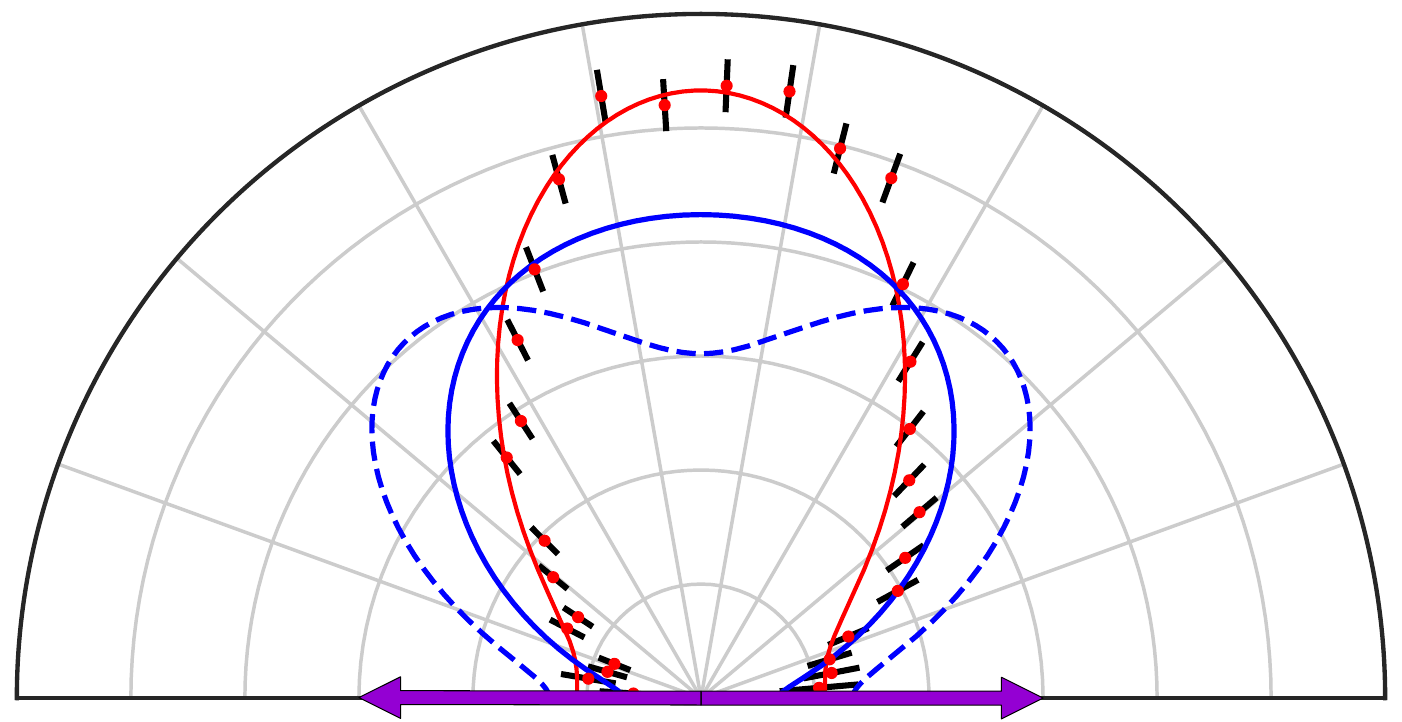}}}
    {
    \subfigure[]{%
        \includegraphics[width=7.0cm, trim= 0.5cm 3.9cm 0.5cm 4cm, clip]{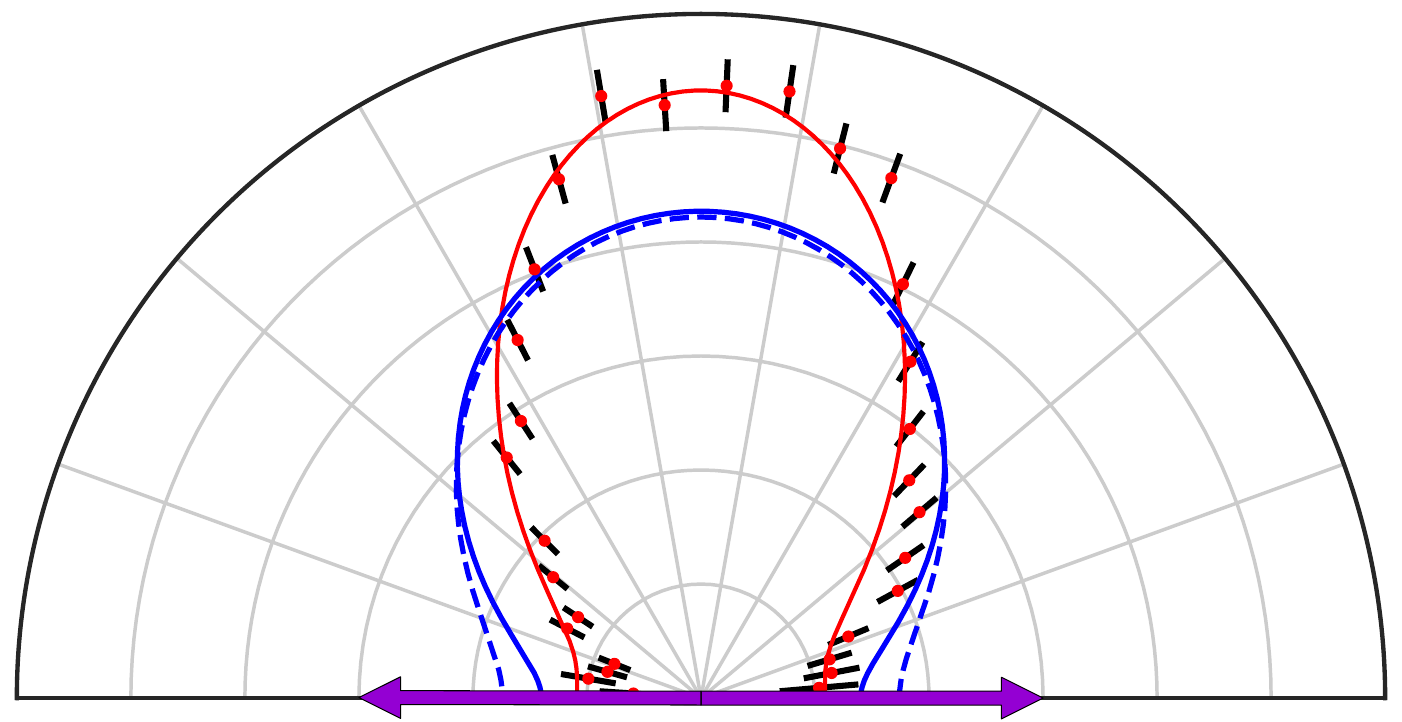}}}
    {
    \subfigure[]{%
        \includegraphics[width=7.0cm, trim= 0.5cm 3.9cm 0.5cm 4cm, clip]{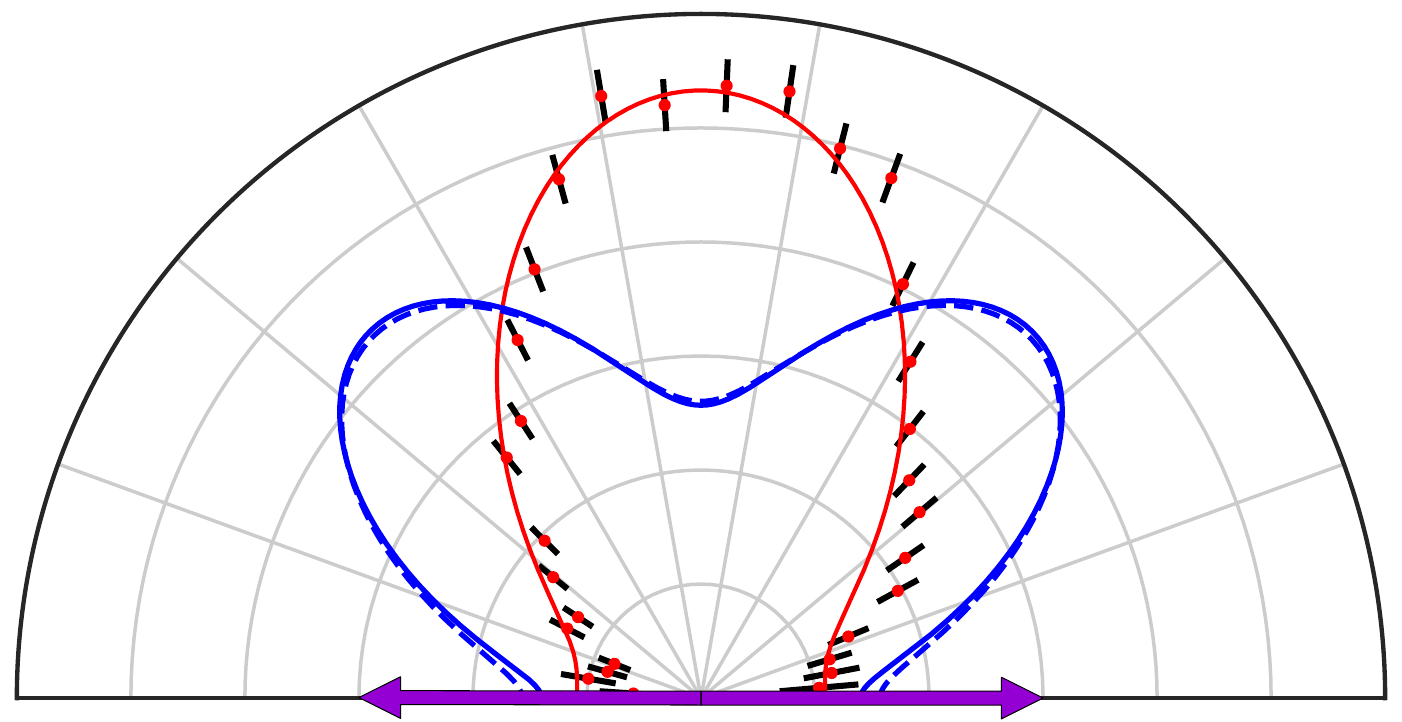}}}
\caption{The photoelectron angular distribution for NOTPSI of Ar at 9.3 eV for the experimentally retrieved $\beta$ parameters (solid red curve in (a), (b), and (c)) and those extracted from \cite{Starace3,Moccia}. The blue curves in (a) correspond with the uncorrelated HF calculation of \cite{Starace3}, the blue curves in (b) correspond with the Coulomb correlated calculation of \cite{Starace3}, and the blue curves in (c) correspond with the random phase approximation calculation of \cite{Moccia} (dashed: velocity gauge, solid: length gauge). The orientation of the VUV polarization is indicated by the horizontal double arrow.}
\label{fig:StaracePAD}
\end{figure}

We compare our retrieved $\beta$ parameters with those extracted from Ref.~\cite{Starace3} at a photon energy of 9.3~eV and Ref.~\cite{Moccia} at a photon energy of 8.6~eV, presented in Table~\ref{table:betas} and Fig.~\ref{fig:StaracePAD}. In Ref. \cite{Starace3}, the calculations were performed using a 2\textsuperscript{nd} order time-independent perturbation theory method, in both a Hartree-Fock (HF) approach and a Coulomb correlated HF approach. In the uncorrelated HF calculation, there is significant disagreement between the length and velocity gauges, while the Coulomb correlated HF approach exhibits much better gauge invariance. We find that the Coulomb correlated HF calculations show good qualitative agreement with the present measurements (seen in Fig.~\ref{fig:StaracePAD} (b)), specifically in the direction of maximum photoelectron emission. There are, however, significant quantitative discrepancies in the magnitude of $\beta_2$ and the sign of $\beta_4$ (see Table~\ref{table:betas}). The uncorrelated HF calculations of Ref.~\cite{Starace3}, in either the length or velocity gauge, compare less favorably with the present measurements (seen in Fig.~\ref{fig:StaracePAD} (a)).

In Ref.~\cite{Moccia}, the two-photon ionization cross sections were calculated using a random phase approximation method with HF wave functions for the initial, intermediate, and target states, neglecting electron-electron correlation. There is reasonable agreement at this photon energy between the calculations in length and velocity gauge (seen in Fig.~\ref{fig:StaracePAD} (c)). They both resemble the uncorrelated results in the velocity gauge of Ref.~\cite{Starace3} shown in Fig.~\ref{fig:StaracePAD}~(a). However, there is poor qualitative and quantitative agreement between the uncorrelated theories and the measurement. Despite the quantitative disagreements, the correlated HF results of Ref.~\cite{Starace3} suggest that electron-electron correlation is essential in the accurate description of the NOTPSI dynamics of argon.

\begin{table}[h!] 
\centering
\begin{tabular}{  c | c c  } 
 \hline\hline
  & $\beta_2$ & $\beta_4$ \\
 \hline
 HF Length \cite{Starace3} & -0.62 & -0.18 \\
 HF Velocity \cite{Starace3} & -0.13 & -0.48 \\
 Correlated Length \cite{Starace3} & -0.54 & -0.05 \\
 Correlated Velocity \cite{Starace3} & -0.48 & -0.01 \\
 Random Phase Approx. Length \cite{Moccia} & 0.03 & -0.62 \\
 Random Phase Approx. Velocity \cite{Moccia} & 0.04 & -0.58 \\
 Experiment & -0.93 & 0.25 \\ 
 \hline
\end{tabular}
\caption{The $\beta$ parameters extracted from the calculations in \cite{Starace3} at a photon energy of 9.3 eV and in \cite{Moccia} at a photon energy of 8.6 eV, and those retrieved from the present measurement.}
\label{table:betas}
\end{table}

The discrepancies between the theory in Refs.~\cite{Starace3,Moccia} and the present measurements may be attributed to an inadequate treatment of electron-electron correlation in the calculations. The level of correlation accounted for in the Coulomb correlated HF approach of Ref.~\cite{Starace3} led to better gauge invariance and a PAD with greater similarity to the present measurements. This suggests that a higher level of electron-electron correlation must be included for a more accurate description of NOTPSI in argon.


In conclusion, we have reported results on NOTPSI of argon using 3-D momentum imaging and an intense 9.3 eV femtosecond pulse. We find that the observed photoelectron emission pattern can be explained by the interference between the different $p$ and $f$ partial wave components of the photoelectron scattering wave function, which add destructively along the polarization direction of the ionizing VUV field. Our measurements are compared against a previous set of calculations, which reveal that the photoionization dynamics are evidently influenced by electron-electron correlation effects. It appears that the level of electron-electron correlation accounted for in the Coulomb correlated HF calculations in Ref.~\cite{Starace3} is not sufficient to reach complete agreement with the present results. Our measurements can serve as a benchmark for future ab initio theoretical treatments of NOTPSI dynamics in multi-electron systems. A particular challenge may be incorporating continuum-continuum coupling in the calculations, which is expected to be important in reproducing the PAD in non-resonant regions \cite{Bello}. In addition to further development of theoretical methods, there is a clear need for follow-up experiments to investigate the photon energy dependence of electron-electron correlation effects by angle-resolved photoionization of multi-electron atoms and small molecules, using intense VUV, XUV, and soft X-rays, preferably at photon energies where calculated anisotropy parameters are gauge invariant, and correlated and uncorrelated results differ markedly.


Work at Lawrence Berkeley National Laboratory was performed under the auspices of the U.S. Department of Energy under Contract No. DE-AC02-05CH11231, and was supported by the U.S. Department of Energy Office of Basic Energy Sciences, Division of Chemical Sciences, Biosciences and Geosciences. This research used resources of the National Energy Research Scientific Computing Center (NERSC), a U.S. Department of Energy Office of Science User Facilities under Contract No. DE-AC02-05CH11231. We are indebted to the RoentDek Company for long-term support with detector software and hardware. We thank Roger Y. Bello, Robert R. Lucchese, and C. William McCurdy for the many helpful discussions.

\bibliography{ArRefs}

\begin{thebibliography}{40}
\providecommand{\natexlab}[1]{#1}
\providecommand{\url}[1]{\texttt{#1}}
\expandafter\ifx\csname urlstyle\endcsname\relax
  \providecommand{\doi}[1]{doi: #1}\else
  \providecommand{\doi}{doi: \begingroup \urlstyle{rm}\Url}\fi

\bibitem[Boll et~al.(2019)Boll, Foj\'on, McCurdy, and Palacios]{Boll}
Diego I.~R. Boll, Omar~A. Foj\'on, C.~W. McCurdy, and Alicia Palacios.
\newblock Angularly resolved two-photon above-threshold ionization of helium.
\newblock \emph{Phys. Rev. A}, 99:\penalty0 023416, Feb 2019.
\newblock \doi{10.1103/PhysRevA.99.023416}.
\newblock URL \url{https://link.aps.org/doi/10.1103/PhysRevA.99.023416}.

\bibitem[Yip et~al.(2013)Yip, Rescigno, McCurdy, and Mart\'{\i}n]{Yip}
F.~L. Yip, T.~N. Rescigno, C.~W. McCurdy, and F.~Mart\'{\i}n.
\newblock Fully differential single-photon double ionization of neon and argon.
\newblock \emph{Phys. Rev. Lett.}, 110:\penalty0 173001, Apr 2013.
\newblock \doi{10.1103/PhysRevLett.110.173001}.
\newblock URL \url{https://link.aps.org/doi/10.1103/PhysRevLett.110.173001}.

\bibitem[Lucchese(1990)]{Lucchese0}
Robert~R. Lucchese.
\newblock Effects of interchannel coupling on the photoionization cross
  sections of carbon dioxide.
\newblock \emph{The Journal of Chemical Physics}, 92\penalty0 (7):\penalty0
  4203--4211, 1990.
\newblock \doi{10.1063/1.457778}.
\newblock URL \url{https://doi.org/10.1063/1.457778}.

\bibitem[Basden and Lucchese(1988)]{Lucchese1}
Bryan Basden and Robert~R. Lucchese.
\newblock Vibrationally resolved cross sections and asymmetry parameters for
  the photoionization of ${\mathrm{n}}_{2}$ with coupling between the
  (3${\ensuremath{\sigma}}_{g}$${)}^{\mathrm{\ensuremath{-}}1}$ and the
  (2${\ensuremath{\sigma}}_{u}$${)}^{\mathrm{\ensuremath{-}}1}$ channels.
\newblock \emph{Phys. Rev. A}, 37:\penalty0 89--97, Jan 1988.
\newblock \doi{10.1103/PhysRevA.37.89}.
\newblock URL \url{https://link.aps.org/doi/10.1103/PhysRevA.37.89}.

\bibitem[Jose et~al.(2014)Jose, Lucchese, and Rescigno]{Lucchese2}
J.~Jose, R.~R. Lucchese, and T.~N. Rescigno.
\newblock Interchannel coupling effects in the valence photoionization of sf6.
\newblock \emph{The Journal of Chemical Physics}, 140\penalty0 (20):\penalty0
  204305, 2014.
\newblock \doi{10.1063/1.4876576}.
\newblock URL \url{https://doi.org/10.1063/1.4876576}.

\bibitem[Stratmann and Lucchese(1992)]{Lucchese3}
R.~E. Stratmann and Robert~R. Lucchese.
\newblock Resonances and the effects of interchannel coupling in the
  photoionization of cs2.
\newblock \emph{The Journal of Chemical Physics}, 97\penalty0 (9):\penalty0
  6384--6395, 1992.
\newblock \doi{10.1063/1.463699}.
\newblock URL \url{https://doi.org/10.1063/1.463699}.

\bibitem[Rescigno et~al.(1993)Rescigno, Lengsfield, and Orel]{Rescigno}
T.~N. Rescigno, B.~H. Lengsfield, and A.~E. Orel.
\newblock Interchannel coupling and ground state correlation effects in the
  photoionization of co.
\newblock \emph{The Journal of Chemical Physics}, 99\penalty0 (7):\penalty0
  5097--5103, 1993.
\newblock \doi{10.1063/1.466010}.
\newblock URL \url{https://doi.org/10.1063/1.466010}.

\bibitem[Houlgate et~al.(1974)Houlgate, Codling, Marr, and West]{Codling}
R~G Houlgate, K~Codling, G~V Marr, and J~B West.
\newblock Angular distribution and photoionization cross section measurements
  on the 3p and 3s subshells of argon.
\newblock \emph{Journal of Physics B: Atomic and Molecular Physics}, 7\penalty0
  (17):\penalty0 L470--L473, dec 1974.
\newblock \doi{10.1088/0022-3700/7/17/003}.
\newblock URL \url{https://doi.org/10.1088%2F0022-3700%2F7%2F17%2F003}.

\bibitem[Altun and Manson(2000)]{Altun}
Zikri Altun and Steven~T. Manson.
\newblock Photoelectron angular distributions of $\mathrm{ns}$ subshells of
  open-shell atoms as indicators of interchannel coupling: $\mathrm{Sc} 4s$
  photoionization.
\newblock \emph{Phys. Rev. A}, 61:\penalty0 030702, Feb 2000.
\newblock \doi{10.1103/PhysRevA.61.030702}.
\newblock URL \url{https://link.aps.org/doi/10.1103/PhysRevA.61.030702}.

\bibitem[Southworth et~al.(1986)Southworth, Parr, Hardis, and Dehmer]{Dehmer}
S.~H. Southworth, A.~C. Parr, J.~E. Hardis, and J.~L. Dehmer.
\newblock Channel coupling and shape resonance effects in the photoelectron
  angular distributions of the
  3${\ensuremath{\sigma}}_{g}^{\mathrm{\ensuremath{-}}1}$ and
  2${\ensuremath{\sigma}}_{u}^{\mathrm{\ensuremath{-}}1}$ channels of
  ${\mathrm{n}}_{2}$.
\newblock \emph{Phys. Rev. A}, 33:\penalty0 1020--1023, Feb 1986.
\newblock \doi{10.1103/PhysRevA.33.1020}.
\newblock URL \url{https://link.aps.org/doi/10.1103/PhysRevA.33.1020}.

\bibitem[Lin(1974)]{Lin}
C.~D. Lin.
\newblock Channel interaction and threshold behavior of photoionization.
\newblock \emph{Phys. Rev. A}, 9:\penalty0 171--180, Jan 1974.
\newblock \doi{10.1103/PhysRevA.9.171}.
\newblock URL \url{https://link.aps.org/doi/10.1103/PhysRevA.9.171}.

\bibitem[Kennedy and Manson(1972)]{Manson}
David~J. Kennedy and Steven~Trent Manson.
\newblock Photoionization of the noble gases: Cross sections and angular
  distributions.
\newblock \emph{Phys. Rev. A}, 5:\penalty0 227--247, Jan 1972.
\newblock \doi{10.1103/PhysRevA.5.227}.
\newblock URL \url{https://link.aps.org/doi/10.1103/PhysRevA.5.227}.

\bibitem[Haber et~al.(2009)Haber, Doughty, and Leone]{Leone}
Louis~H Haber, Benjamin Doughty, and Stephen~R Leone.
\newblock Photoelectron angular distributions and cross section ratios of
  two-color two-photon above threshold ionization of argon.
\newblock \emph{The Journal of Physical Chemistry A}, 113\penalty0
  (47):\penalty0 13152--13158, 2009.

\bibitem[Mondal et~al.(2014)Mondal, Fukuzawa, Motomura, Tachibana, Nagaya,
  Sakai, Matsunami, Yase, Yao, Wada, Hayashita, Saito, Callegari, Prince,
  Miron, Nagasono, Togashi, Yabashi, Ishikawa, Kazansky, Kabachnik, and
  Ueda]{Mondal}
S.~Mondal, H.~Fukuzawa, K.~Motomura, T.~Tachibana, K.~Nagaya, T.~Sakai,
  K.~Matsunami, S.~Yase, M.~Yao, S.~Wada, H.~Hayashita, N.~Saito, C.~Callegari,
  K.~C. Prince, C.~Miron, M.~Nagasono, T.~Togashi, M.~Yabashi, K.~L. Ishikawa,
  A.~K. Kazansky, N.~M. Kabachnik, and K.~Ueda.
\newblock Pulse-delay effects in the angular distribution of near-threshold euv
  + ir two-photon ionization of ne.
\newblock \emph{Phys. Rev. A}, 89:\penalty0 013415, Jan 2014.
\newblock \doi{10.1103/PhysRevA.89.013415}.
\newblock URL \url{https://link.aps.org/doi/10.1103/PhysRevA.89.013415}.

\bibitem[D{\"u}sterer et~al.(2019)D{\"u}sterer, Hartmann, Bomme, Boll,
  Costello, Erk, De~Fanis, Ilchen, Johnsson, Kelly, et~al.]{Dusterer}
Stefan D{\"u}sterer, G~Hartmann, C~Bomme, R~Boll, JT~Costello, B~Erk,
  A~De~Fanis, M~Ilchen, P~Johnsson, TJ~Kelly, et~al.
\newblock Two-color xuv+ nir femtosecond photoionization of neon in the
  near-threshold region.
\newblock \emph{New Journal of Physics}, 21\penalty0 (6):\penalty0 063034,
  2019.

\bibitem[Miyamoto et~al.(2004)Miyamoto, Kamei, Yoshitomi, Kanai, Sekikawa,
  Nakajima, and Watanabe]{Miyamoto}
Naoki Miyamoto, Masato Kamei, Dai Yoshitomi, Teruto Kanai, Taro Sekikawa,
  Takashi Nakajima, and Shuntaro Watanabe.
\newblock Observation of two-photon above-threshold ionization of rare gases by
  xuv harmonic photons.
\newblock \emph{Phys. Rev. Lett.}, 93:\penalty0 083903, Aug 2004.
\newblock \doi{10.1103/PhysRevLett.93.083903}.
\newblock URL \url{https://link.aps.org/doi/10.1103/PhysRevLett.93.083903}.

\bibitem[Sekikawa et~al.(2004)Sekikawa, Kosuge, Kanai, and Watanabe]{Sekikawa3}
Taro Sekikawa, Atsushi Kosuge, Teruto Kanai, and Shuntaro Watanabe.
\newblock Nonlinear optics in the extreme ultraviolet.
\newblock \emph{Nature}, 432\penalty0 (7017):\penalty0 605--608, 2004.

\bibitem[Sato et~al.(2011)Sato, Iwasaki, Ishibashi, Okino, Yamanouchi, Adachi,
  Yagishita, Yazawa, Kannari, Aoyma, et~al.]{Sato}
Takahiro Sato, Atsushi Iwasaki, Kazuki Ishibashi, Tomoya Okino, Kaoru
  Yamanouchi, Junichi Adachi, Akira Yagishita, Hiroki Yazawa, Fumihiko Kannari,
  Makoto Aoyma, et~al.
\newblock Determination of the absolute two-photon ionization cross section of
  he by an xuv free electron laser.
\newblock \emph{Journal of Physics B: Atomic, Molecular and Optical Physics},
  44\penalty0 (16):\penalty0 161001, 2011.

\bibitem[Meyer et~al.(2010)Meyer, Cubaynes, Richardson, Costello, Radcliffe,
  Li, D\"usterer, Fritzsche, Mihelic, Papamihail, and Lambropoulos]{Meyer}
M.~Meyer, D.~Cubaynes, V.~Richardson, J.~T. Costello, P.~Radcliffe, W.~B. Li,
  S.~D\"usterer, S.~Fritzsche, A.~Mihelic, K.~G. Papamihail, and
  P.~Lambropoulos.
\newblock Two-photon excitation and relaxation of the
  $3d\mathrm{\text{\ensuremath{-}}}4d$ resonance in atomic kr.
\newblock \emph{Phys. Rev. Lett.}, 104:\penalty0 213001, May 2010.
\newblock \doi{10.1103/PhysRevLett.104.213001}.
\newblock URL \url{https://link.aps.org/doi/10.1103/PhysRevLett.104.213001}.

\bibitem[Ma et~al.(2013)Ma, Motomura, Ishikawa, Mondal, Fukuzawa, Yamada, Ueda,
  Nagaya, Yase, Mizoguchi, et~al.]{Ma}
R~Ma, K~Motomura, KL~Ishikawa, S~Mondal, H~Fukuzawa, A~Yamada, K~Ueda,
  K~Nagaya, S~Yase, Y~Mizoguchi, et~al.
\newblock Photoelectron angular distributions for the two-photon ionization of
  helium by ultrashort extreme ultraviolet free-electron laser pulses.
\newblock \emph{Journal of Physics B: Atomic, Molecular and Optical Physics},
  46\penalty0 (16):\penalty0 164018, 2013.

\bibitem[Pan and Starace(1991)]{Starace3}
Cheng Pan and Anthony~F. Starace.
\newblock Angular distribution of electrons following two-photon ionization of
  the ar atom and two-photon detachment of the
  ${\mathrm{f}}^{\mathrm{\ensuremath{-}}}$ ion.
\newblock \emph{Phys. Rev. A}, 44:\penalty0 324--329, Jul 1991.
\newblock \doi{10.1103/PhysRevA.44.324}.
\newblock URL \url{https://link.aps.org/doi/10.1103/PhysRevA.44.324}.

\bibitem[Moccia et~al.(1983)Moccia, Rahman, and Rizzo]{Moccia}
R~Moccia, N~K Rahman, and A~Rizzo.
\newblock Two-photon ionisation cross section calculations of noble gases:
  results for ne and ar.
\newblock \emph{Journal of Physics B: Atomic and Molecular Physics},
  16\penalty0 (15):\penalty0 2737--2751, aug 1983.
\newblock \doi{10.1088/0022-3700/16/15/016}.
\newblock URL \url{https://doi.org/10.1088%2F0022-3700%2F16%2F15%2F016}.

\bibitem[Dörner et~al.(2000)Dörner, Mergel, Jagutzki, Spielberger, Ullrich,
  Moshammer, and Schmidt-Böcking]{dornerColdTargetRecoil2000}
R.~Dörner, V.~Mergel, O.~Jagutzki, L.~Spielberger, J.~Ullrich, R.~Moshammer,
  and H.~Schmidt-Böcking.
\newblock Cold {Target} {Recoil} {Ion} {Momentum} {Spectroscopy}: a 'momentum
  microscope' to view atomic collision dynamics.
\newblock \emph{Physics Reports}, 330\penalty0 (2-3):\penalty0 95--192, June
  2000.
\newblock ISSN 0370-1573.
\newblock \doi{doi: DOI: 10.1016/S0370-1573(99)00109-X}.
\newblock URL
  \url{http://www.sciencedirect.com/science/article/B6TVP-401HH57-1/2/587a27ccbfe492bcbe5b72191579ddbd}.

\bibitem[Ullrich et~al.(2003)Ullrich, Moshammer, Dorn, D{\"o}rner, Schmidt, and
  Schmidt-B{\"o}cking]{Ullrich}
Joachim Ullrich, Robert Moshammer, Alexander Dorn, Reinhard D{\"o}rner, L~Ph~H
  Schmidt, and H~Schmidt-B{\"o}cking.
\newblock Recoil-ion and electron momentum spectroscopy: reaction-microscopes.
\newblock \emph{Reports on Progress in Physics}, 66\penalty0 (9):\penalty0
  1463, 2003.

\bibitem[Jahnke et~al.(2004)Jahnke, Weber, Osipov, Landers, Jagutzki, Schmidt,
  Cocke, Prior, Schmidt-B{\"o}cking, and D{\"o}rner]{Jahnke}
T~Jahnke, Th~Weber, T~Osipov, AL~Landers, O~Jagutzki, L~Ph~H Schmidt, CL~Cocke,
  MH~Prior, H~Schmidt-B{\"o}cking, and R~D{\"o}rner.
\newblock Multicoincidence studies of photo and auger electrons from
  fixed-in-space molecules using the coltrims technique.
\newblock \emph{Journal of Electron Spectroscopy and Related Phenomena},
  141\penalty0 (2-3):\penalty0 229--238, 2004.

\bibitem[Sturm et~al.(2016)Sturm, Wright, Ray, Zalyubovskaya, Shivaram,
  Slaughter, Ranitovic, Belkacem, and Weber]{Sturm}
F.~P. Sturm, T.~W. Wright, D.~Ray, I.~Zalyubovskaya, N.~Shivaram, D.~S.
  Slaughter, P.~Ranitovic, A.~Belkacem, and Th. Weber.
\newblock Time resolved 3d momentum imaging of ultrafast dynamics by coherent
  vuv-xuv radiation.
\newblock \emph{Review of Scientific Instruments}, 87\penalty0 (6):\penalty0
  063110, 2016.
\newblock \doi{10.1063/1.4953441}.
\newblock URL \url{https://doi.org/10.1063/1.4953441}.

\bibitem[Roentdek()]{Roentdek}
Roentdek.
\newblock Roentdek delayline detectors.
\newblock URL \url{http://www.roentdek.com}.

\bibitem[Jagutzki et~al.(2002)Jagutzki, Cerezo, Czasch, Dorner, Hattas, Huang,
  Mergel, Spillmann, Ullmann-Pfleger, Weber, et~al.]{Jagutzki}
Ottmar Jagutzki, Alfred Cerezo, Achim Czasch, R~Dorner, M~Hattas, Min Huang,
  Volker Mergel, Uwe Spillmann, Klaus Ullmann-Pfleger, Thorsten Weber, et~al.
\newblock Multiple hit readout of a microchannel plate detector with a
  three-layer delay-line anode.
\newblock \emph{IEEE Transactions on Nuclear Science}, 49\penalty0
  (5):\penalty0 2477--2483, 2002.

\bibitem[Allison et~al.(2009)Allison, van Tilborg, Wright, Hertlein, Falcone,
  and Belkacem]{Allison}
T.~K. Allison, J.~van Tilborg, T.~W. Wright, M.~P. Hertlein, R.~W. Falcone, and
  A.~Belkacem.
\newblock Separation of high order harmonics with fluoride windows.
\newblock \emph{Opt. Express}, 17\penalty0 (11):\penalty0 8941--8946, May 2009.
\newblock \doi{10.1364/OE.17.008941}.
\newblock URL
  \url{http://www.opticsexpress.org/abstract.cfm?URI=oe-17-11-8941}.

\bibitem[Li(1980)]{Li}
H.~H. Li.
\newblock Refractive index of alkaline earth halides and its wavelength and
  temperature derivatives.
\newblock \emph{Journal of Physical and Chemical Reference Data}, 9\penalty0
  (1):\penalty0 161--290, 1980.
\newblock \doi{10.1063/1.555616}.
\newblock URL \url{https://doi.org/10.1063/1.555616}.

\bibitem[Sekikawa et~al.(2002)Sekikawa, Katsura, Miura, and
  Watanabe]{Sekikawa1}
Taro Sekikawa, Tomotaka Katsura, Satoshi Miura, and Shuntaro Watanabe.
\newblock Measurement of the intensity-dependent atomic dipole phase of a high
  harmonic by frequency-resolved optical gating.
\newblock \emph{Phys. Rev. Lett.}, 88:\penalty0 193902, Apr 2002.
\newblock \doi{10.1103/PhysRevLett.88.193902}.
\newblock URL \url{https://link.aps.org/doi/10.1103/PhysRevLett.88.193902}.

\bibitem[Sekikawa et~al.(1999)Sekikawa, Ohno, Yamazaki, Nabekawa, and
  Watanabe]{Sekikawa2}
Taro Sekikawa, Tomoki Ohno, Tomohiro Yamazaki, Yasuo Nabekawa, and Shuntaro
  Watanabe.
\newblock Pulse compression of a high-order harmonic by compensating the atomic
  dipole phase.
\newblock \emph{Phys. Rev. Lett.}, 83:\penalty0 2564--2567, Sep 1999.
\newblock \doi{10.1103/PhysRevLett.83.2564}.
\newblock URL \url{https://link.aps.org/doi/10.1103/PhysRevLett.83.2564}.

\bibitem[Larsen et~al.(2016)Larsen, Cryan, Shivaram, Champenois, Wright, Ray,
  Kostko, Ahmed, Belkacem, and Slaughter]{Larsen}
K.~A. Larsen, J.~P. Cryan, N.~Shivaram, E.~G. Champenois, T.~W. Wright, D.~Ray,
  O.~Kostko, M.~Ahmed, A.~Belkacem, and D.~S. Slaughter.
\newblock Vuv and xuv reflectance of optically coated mirrors for selection of
  high harmonics.
\newblock \emph{Opt. Express}, 24\penalty0 (16):\penalty0 18209--18216, Aug
  2016.
\newblock \doi{10.1364/OE.24.018209}.
\newblock URL
  \url{http://www.opticsexpress.org/abstract.cfm?URI=oe-24-16-18209}.

\bibitem[Reid(2003)]{Reid}
Katharine~L Reid.
\newblock Photoelectron angular distributions.
\newblock \emph{Annual review of physical chemistry}, 54\penalty0 (1):\penalty0
  397--424, 2003.

\bibitem[Liu et~al.(2007)Liu, Lucchese, Grum-Grzhimailo, Morishita, Saito,
  Prümper, and Ueda]{Liu}
X-J Liu, R~R Lucchese, A~N Grum-Grzhimailo, Y~Morishita, N~Saito, G~Prümper,
  and K~Ueda.
\newblock Molecular-frame photoelectron and electron-frame photoion angular
  distributions and their interrelation.
\newblock \emph{Journal of Physics B: Atomic, Molecular and Optical Physics},
  40\penalty0 (3):\penalty0 485--496, jan 2007.
\newblock \doi{10.1088/0953-4075/40/3/004}.
\newblock URL \url{https://doi.org/10.1088%2F0953-4075%2F40%2F3%2F004}.

\bibitem[Efron(1992)]{Efron}
Bradley Efron.
\newblock Bootstrap methods: another look at the jackknife.
\newblock In \emph{Breakthroughs in statistics}, pages 569--593. Springer,
  1992.

\bibitem[Mabbs et~al.(2009)Mabbs, Grumbling, Pichugin, and Sanov]{Mabbs}
Richard Mabbs, Emily~R. Grumbling, Kostyantyn Pichugin, and Andrei Sanov.
\newblock Photoelectron imaging: an experimental window into electronic
  structure.
\newblock \emph{Chem. Soc. Rev.}, 38:\penalty0 2169--2177, 2009.
\newblock \doi{10.1039/B815748K}.
\newblock URL \url{http://dx.doi.org/10.1039/B815748K}.

\bibitem[Hall and Siegel(1968)]{Hall}
J.~L. Hall and M.~W. Siegel.
\newblock Angular dependence of the laser photodetachment of the negative ions
  of carbon, oxygen, and hydrogen.
\newblock \emph{The Journal of Chemical Physics}, 48\penalty0 (2):\penalty0
  943--945, 1968.
\newblock \doi{10.1063/1.1668743}.
\newblock URL \url{https://doi.org/10.1063/1.1668743}.

\bibitem[Cooper and Zare(1968)]{Zare}
J.~Cooper and R.~N. Zare.
\newblock Angular distribution of photoelectrons.
\newblock \emph{The Journal of Chemical Physics}, 48\penalty0 (2):\penalty0
  942--943, 1968.
\newblock \doi{10.1063/1.1668742}.
\newblock URL \url{https://doi.org/10.1063/1.1668742}.

\bibitem[Bello(2020)]{Bello}
Roger~Y. Bello.
\newblock private communication, Mar 2020.

\end{thebibliography}

\end{document}